\documentclass[12pt]{arXiv_class}

\title{\textbf{Complementarity Analysis of Interference between Frequency-Displaced Photonic Wave-Packets}} 
\author{Gustavo C. Amaral, Elisa F. Carneiro, Guilherme P. Tempor\~{a}o \\and Jean Pierre von der Weid\\\textit{Corresponding author: gustavo@opto.cetuc.puc-rio.br}}

\begin{document}
\maketitle

\begin{abstract}
The complementarity relation between the visibility and the spectral distinguishability of frequency-displaced photonic wave-packets in a Hong-Ou-Mandel interferometer is studied. An experimental definition of $K$, the distinguishability parameter, is proposed and tested for the $K^2+\mathcal{V}^2\leq 1$ complementarity inequality when a consistent visibility parameter is defined. The results show that the spectral distinguishability is, indeed, complementary to the visibility and that the quantum aspect of the two-photon interference phenomenon can be examined by employing weak-coherent states.
\end{abstract}

In 1984, Hong, Ou, and Mandel developed an experiment capable of quantifying the degree of distinguishability between two photonic quantum states, the Hong-Ou-Mandel interferometer \cite{hong1987measurement}. Two indistinguishable single-photons entering a symmetrical beam-splitter from different input ports are incapable of leaving the device through different output arms. The two-photon wave-packet that describes the collective input state experiences destructive interference and the photons leave the beam-splitter ``bunched" together. The phenomenon is a fundamentally quantum one which translates the degree of distinguishability between the individual input states directly as the visibility of the Hong-Ou-Mandel interferogram: unitary visibility meaning complete indistinguishability; and null visibility meaning complete distinguishability. With the advent of quantum memories, the HOM interference has attracted great attention: a quantum memory's ability to preserve the entire photonic wave-packet can be assessed by measuring the visibility of the HOM interferogram after the states are stored and recovered from it \cite{jin2013two}. Also, the HOM interference phenomenon is at the heart of the projection onto the Bell state basis \cite{lo2012measurement}.

HOM interference and indistinguishability between photonic wave-packets gained great interest with the development of the Measurement-Device Independent Quantum Key Distribution (MDI-QKD) protocol \cite{lo2012measurement}. Single-photons, which would be ideal for QKD protocols, are scarcely available, and faint laser pulses, or weak-coherent state (WCS) pulses, are employed as an approximation of a single-photon state \cite{gisin2002quantum}. HOM interference with WCSs, however, cannot reach unitary visibility even with perfect indistinguishability due to a non-zero multi-photon emission probability, with a maximum achievable visibility limited to $50\%$ \cite{mandel1983photon}. Nevertheless, and due to the fact that a WCS pulse will probabilistically contain a single photon, two-photon HOM interference of WCSs has been explained by a statistical decomposition of the input pairs of WCS pulses into pairs of Fock states; each possible outcome being weighted by its respective probability of occurrence \cite{da2015linear}.

In a single-mode optical fiber setup, the spatial mode is pre-determined so that the degrees of freedom that may distinguish two photonic wave-packets are: their polarization mode; the mean number of photons; their temporal modes; and their spectral modes. In fact, a recent result has been presented where, by guaranteeing the indistinguishability of all degrees of freedom except for the spectral mode, the latter can be determined by analyzing the HOM interference of the wave-packets \cite{amaral2016few}. This result raises a question regarding the very nature of the HOM interference: how can two completely distinguishable photonic wave-packets (the spectral modes are disjoint) produce non-null HOM interference when the phenomenon is, itself, dependent on their indistinguishability?

Here, the proposition that the interference phenomenon is a result of an impossibility of the detectors to identify the individual spectral distribution of the input states is presented. For that, an experimental definition of the spectral distinguishability parameter between two photonic wave-packets is presented and shown to obey a complementarity relation when WCSs are employed. The impact of the presented results is two-fold: first, it shows that the spectral distinguishability is complementary to the visibility of the HOM interferogram; and, second, shows that the two-photon interference phenomenon can be distilled from the WCS interference in a HOM interferometer, since complementarity is a strictly quantum characteristic.

\begin{figure}[htbp]
\centering
\fbox{\includegraphics[width=0.77\linewidth]{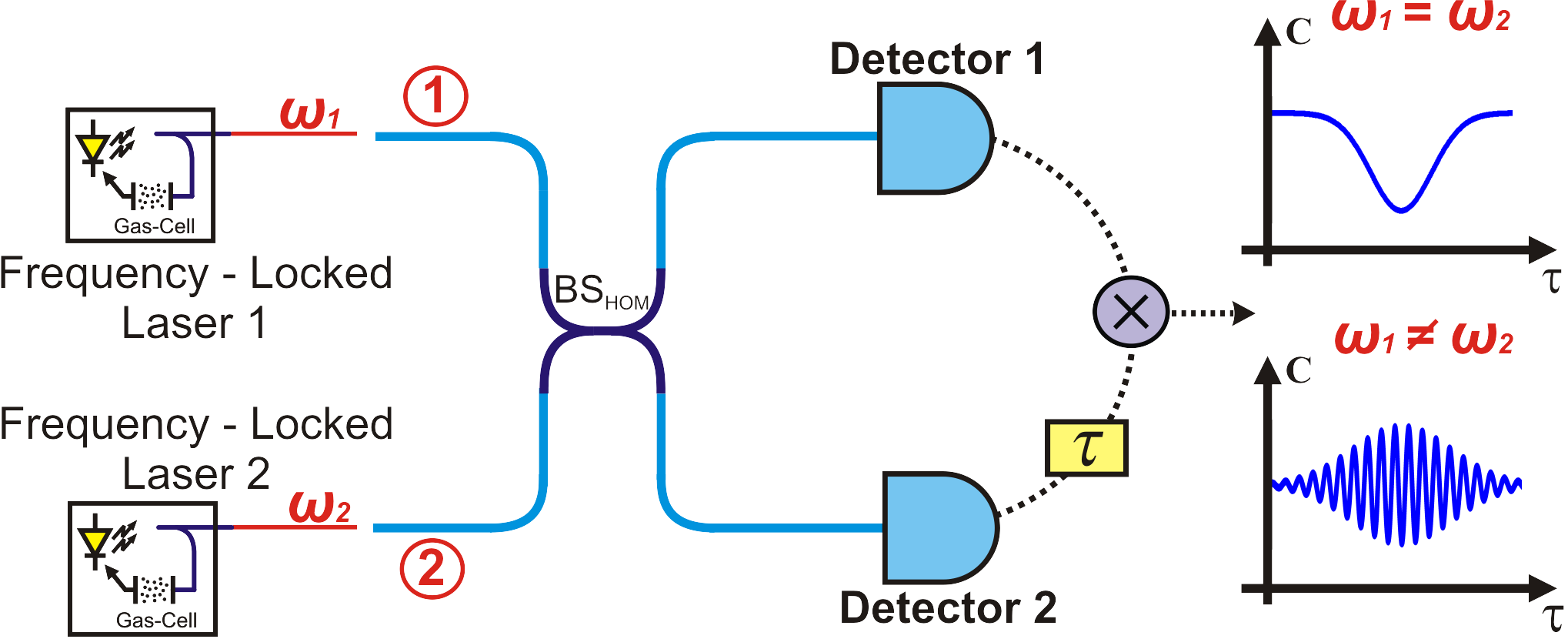}}
\caption{Time-resolved HOM interference setup: when the optical center frequencies $\omega_1$ and $\omega_2$ are the same, the result is the usual HOM dip; when $\omega_1\neq \omega_2$, the beat note is translated in the interferogram as interference fringes \cite{amaral2016few}.}
\label{fig:timeResHOMInt_prop}
\end{figure}

The setup employed to examine the HOM interference of frequency-displaced photonic wave-packets is simple, and is depicted in Fig. \ref{fig:timeResHOMInt_prop}: two optical beams with identical polarization modes, spatial modes, and mean number of photons but centered at different frequencies, are sent to a time-resolved HOM interferometer; by adjusting the relative time $\tau$ between detections, the temporal mode of the wave-packets is synchronized. If the wave-packets have identical frequency modes, the result of the interferogram, as $\tau$ is swept, is the usual HOM dip \cite{da2013proof}; however, in the case the center frequencies are different, the beat note between them is observed in the interferogram \cite{amaral2016few, legero2003time}.

An important observation is that, in the configuration of Fig. \ref{fig:timeResHOMInt_prop}, the spatio-temporal modes and, thus, the frequency modes, are defined by the optical path, i.e., the wave-packet with frequency mode centered at $\omega_{1,2}$ comes from input path $1,2$. Any two single-photons generated by the optical sources that are directed to the interferometer through its upper (path \textbf{1}) or lower (path \textbf{2)} arms will be respectively described as
\begin{align}
\begin{aligned}
\left| 1 \right>_1 &= \frac{e^{-\left(t-\tau_1\right)^2/\left(2\sigma_1^2\right)}}{\sigma_1\sqrt{2\pi}}e^{-i\left(\omega_1\right)}\hat{a}^{\dagger}_1\left| 0 \right>=f_1\left(t\right)\hat{a}^{\dagger}_1\left| 0 \right>\\
\left| 1 \right>_2 &= \frac{e^{-\left(t-\tau_2\right)^2/\left(2\sigma_2^2\right)}}{\sigma_2\sqrt{2\pi}}e^{-i\left(\omega_2\right)}\hat{a}^{\dagger}_2\left| 0 \right>=f_2\left(t\right)\hat{a}^{\dagger}_2\left| 0 \right>,
\end{aligned}
\label{eq:spatioTemporal}
\end{align}
where the fact that the sources emit photons with Gaussian-shaped wave-packets in two well-defined frequency modes $\omega_1$ and $\omega_2$, have been assumed for simplicity. $\tau_{1,2}$ represent the relative delays of the wave-packets that must be compensated in the time-resolved HOM interferometer; $\sigma_{1,2}$ represent the half-width at $1/e$ of the wave-packets; and $\hat{a}^{\dagger}_{1,2}$ are the creation operators for spatial modes $1$ and $2$. The position of the beam splitter has been taken as a reference, so all spatial dependence of $f_{1,2}\left(t\right)$ has been neglected \cite{legero2003time}.

In the particular case where the wave-packets are monochromatic (i.e., their spectral distributions are unit impulses), and the integration time of the detectors is sufficiently long, two wave-packets of different frequencies will always be distinguishable. In practice, however, the detectors have a frequency response and the linewidth of the wave-packets is different from zero, which leaves margin for errors in distinguishing the provenance of the photons; inside the region of indetermination where the spectral distributions overlap and the provenance of the photons cannot be perfectly determined, the wave-packets are indistinguishable and produce an interference pattern even though frequency-displaced. Writing the spectral decomposition of the interfering wave-packets, one has:
\begin{align}
\begin{aligned}
\left.|1\right>_1 &= \int_{-\infty}^{\infty}d\omega X_1\left(\omega\right)\hat{a}^{\dagger}_1\left(\omega\right)\left.|0\right>\\
\left.|1\right>_2 &= \int_{-\infty}^{\infty}d\omega X_2\left(\omega\right)\hat{a}^{\dagger}_2\left(\omega\right)\left.|0\right>,
\end{aligned}
\label{eq:spectralDecomp}
\end{align}
where $X_1\left(\omega\right)$ and $X_2\left(\omega\right)$ are the Fourier transforms of the spatio-temporal functions $f_1\left(t\right)$ and $f_2\left(t\right)$ of Eq. \ref{eq:spatioTemporal}, respectively. Based on Eq. \ref{eq:spectralDecomp}, it is interesting to develop a physical notion of the region of indistinguishability between the wave-packets, i.e., the region within which the two-photon HOM interference will take place; a natural means of doing so is through the \textit{fidelity} between these quantum states.

The fidelity measures the probability of \textit{confusing} two quantum states if one is allowed to perform a single measurement over the system and, therefore, translates the distinguishability between them \cite{vedral2006introduction}. If two quantum states, say $\left.|\rho\right>$ and $\left.|\sigma\right>$, are orthogonal (hence, perfectly distinguishable), the fidelity of these states, calculated as $F\left(\left.|\rho\right>,\left.|\sigma\right>\right) = |\left<\rho|\sigma\right>|^2$, equals zero; in case they are completely indistinguishable, $F\left(\left.|\rho\right>,|\left.\sigma\right>\right)=1$. Calculating the \textit{spectral} fidelity between the interfering states is straightforward from the spectral decomposition of the spatio-temporal mode and from Eq. \ref{eq:spectralDecomp}:
\begin{align}
\begin{aligned}
|\!\left<1|1\right>_{1,2}\!|^2\!&=\\
\bigg|\!\int\!\int\!d\omega_1&d\omega_2X_1\!\left(\omega_1\right)\!X_2^{\ast}\left(\omega_2\right)\!\left<0|\right._1\!\hat{a}\left(\omega_1\right)\!\hat{a}^{\dagger}\!\left(\omega_2\right)\!\left.|0\right>_2\!\bigg|^2.
\end{aligned}
\end{align}
Upon close inspection of the above expression, one notes that the inner product on the rightmost part of the expression can be simplified, i.e., $\left<0|\right._1\hat{a}\left(\omega_1\right)\hat{a}^{\dagger}\left(\omega_2\right)\left.|0\right>_2 = \delta\left(\omega_1-\omega_2\right)$. This allows one to rewrite the expression in a simpler form:
\begin{equation}
F = \bigg|\int_{-\infty}^{\infty}d\omega X_1\left(\omega\right)X_2^{\ast}\left(\omega\right)\bigg|^2.
\label{eq:specFid}
\end{equation}

Note that, since the fidelity equals $1$ whenever a measurement cannot distinguish between the states and zero whenever the states are completely distinguishable, a parameter of \textit{distinguishability} between the states, say $K$, would take the form $K=1-F$, where $F$ is the fidelity as calculated in Eq. \ref{eq:specFid}. Furthermore, in order to calculate the fidelity between two interfering states in a time-tuned HOM interferometer, an extra simplification can be imputed into Eq. \ref{eq:specFid}: for $\tau$ such that the temporal modes are perfectly synchronized, i.e., in the center of the HOM dip, the phases are fixed and can be removed from the integral. In fact, the phase difference $\Delta \phi = \phi_1 - \phi_2$ will be zero and, from the interference equation \cite{saleh1991fundamentals}, the cosine factor that appears multiplying the result will be unit. This way, the multiplication of two otherwise complex functions $X_1\left(\omega\right)$ and $X_2^{\ast}\left(\omega\right)$ simplify to the multiplication of the modulus of these complex functions, which is always positive. In other words, the parameter $K$ of distinguishability may be rewritten, in the center of the HOM dip, as:
\begin{equation}
K = 1 - \left|\int_{-\infty}^{\infty}d\omega |X_1\left(\omega\right)| |X_2\left(\omega\right)|\right|^2.
\label{eq:Kfirst}
\end{equation}

In order to measure the spectral distributions $X_1\left(\omega\right)$ and $X_2\left(\omega\right)$, one might resort to one of two techniques depending on the maximum intensity available for the optical sources: if the intensity is in the few-photon regime, the solution is to make use of the Few-Photon Heterodyne Spectroscopy described in \cite{amaral2016few}; if, on the other hand, the sources are attenuated laser sources, a non-attenuated sample of the optical signal may be directed to a classical high-resolution Optical Spectrum Analyser (OSA). Despite the practicality of the second technique, care must be taken due to the nature of the measurement in an OSA: the intensity of the light fields is measured rather than the field itself. Fortunately, the intensity can be related to the electric field by $I_{1,2}\propto\left|X_{1,2}\right|^2$ and $|X_{1,2}|$, the distribution one must determine for Eq. \ref{eq:Kfirst}, and $I_{1,2}$ are all strictly positive. Therefore, using the positiveness of both $|X_{1,2}|$ and $I_{1,2}$, the expression of $K$ may be written as a function of the measurement of a standard OSA as
\begin{equation}
K = 1 - \alpha\left|\int_{-\infty}^{\infty}d\omega \sqrt{I_1\left(\omega\right)I_2\left(\omega\right)}\right|^2,
\label{eq:intensityK}
\end{equation}
where $\alpha$ is a normalization factor since the integral in Eq. \ref{eq:intensityK} may not be normalized. This integral, however, represents the second cross-moment of two distributions, which has a natural way of being normalized: $\alpha$ corresponds to the inverse of the square root of the products of the individual second moments of each distribution \cite{lewis1995fast}. This has the upside of yielding a dimensionless figure of merit, which agrees with the definition of $K$. Substituting $\alpha$ in Eq. \ref{eq:intensityKnorm} leads to the final expression of $K$:
\begin{equation}
K = 1 - \left|\frac{\int_{-\infty}^{\infty}d\omega \sqrt{I_1\left(\omega\right)I_2\left(\omega\right)}}{\sqrt{\int_{-\infty}^{\infty}d\omega I_1\left(\omega\right) \int_{-\infty}^{\infty}d\omega I_2\left(\omega\right)}}\right|^2.
\label{eq:intensityKnorm}
\end{equation}

\begin{figure}[htbp]
\centering
\fbox{\includegraphics[width=0.77\linewidth]{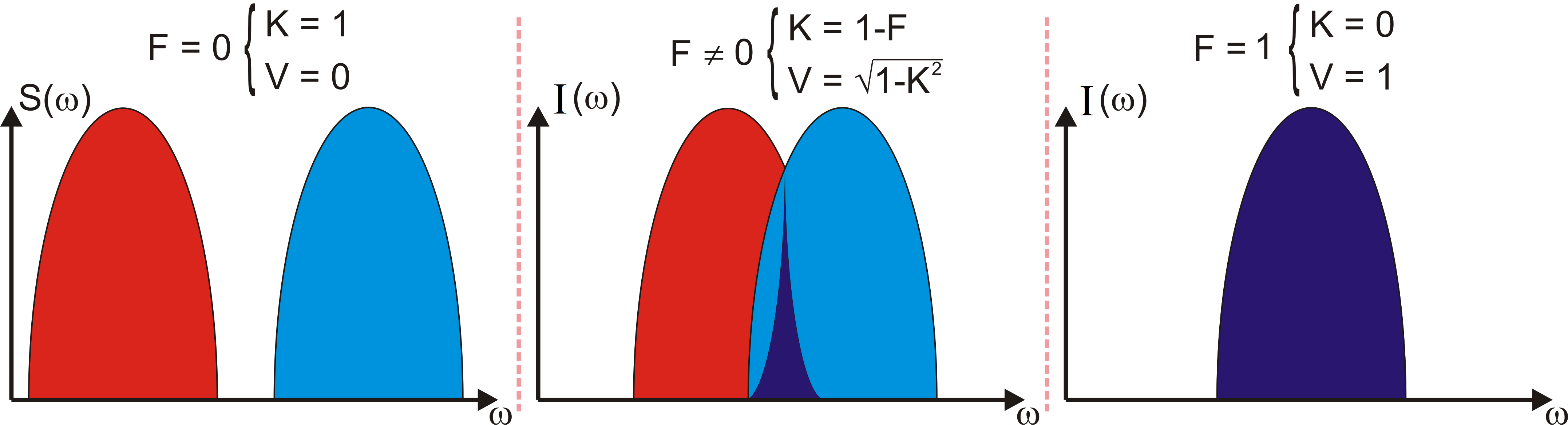}}
\caption{Interpretation of $K$ and $V$ as dependent on the spectral separation of the photonic wave-packets. a) Null intersection between the spectral distributions, $K=1$ and $V=0$; b) non-null intersection, interference fringes can be observed; c) indistinguishable spectra enable maximum visibility, $K=0$ and $V=1$.}
\label{fig:specWhichWay_Kvis}
\end{figure}

Using the presented spectral definition of $K$, the herewith proposition is that the visibility $V$ in a HOM interferometer is tied to $K$ by the complementarity relation $K^2+V^2=1$ \cite{englert1996fringe}. Analogously to \cite{jin2013two, da2015linear}, the visibility is defined as $V=\left(R_{dist}-R_{min}\right)/R_{dist}$, where $R_{dist}$ is the count rate outside the interval of mutual coherence of the wave-packets. In Fig. \ref{fig:specWhichWay_Kvis},a graphical interpretation of the proposed complementarity relation between the spectral measure of $K$ and $V$ is presented for clarity.

\begin{figure}[htbp]
\centering
\fbox{\includegraphics[width=0.77\linewidth]{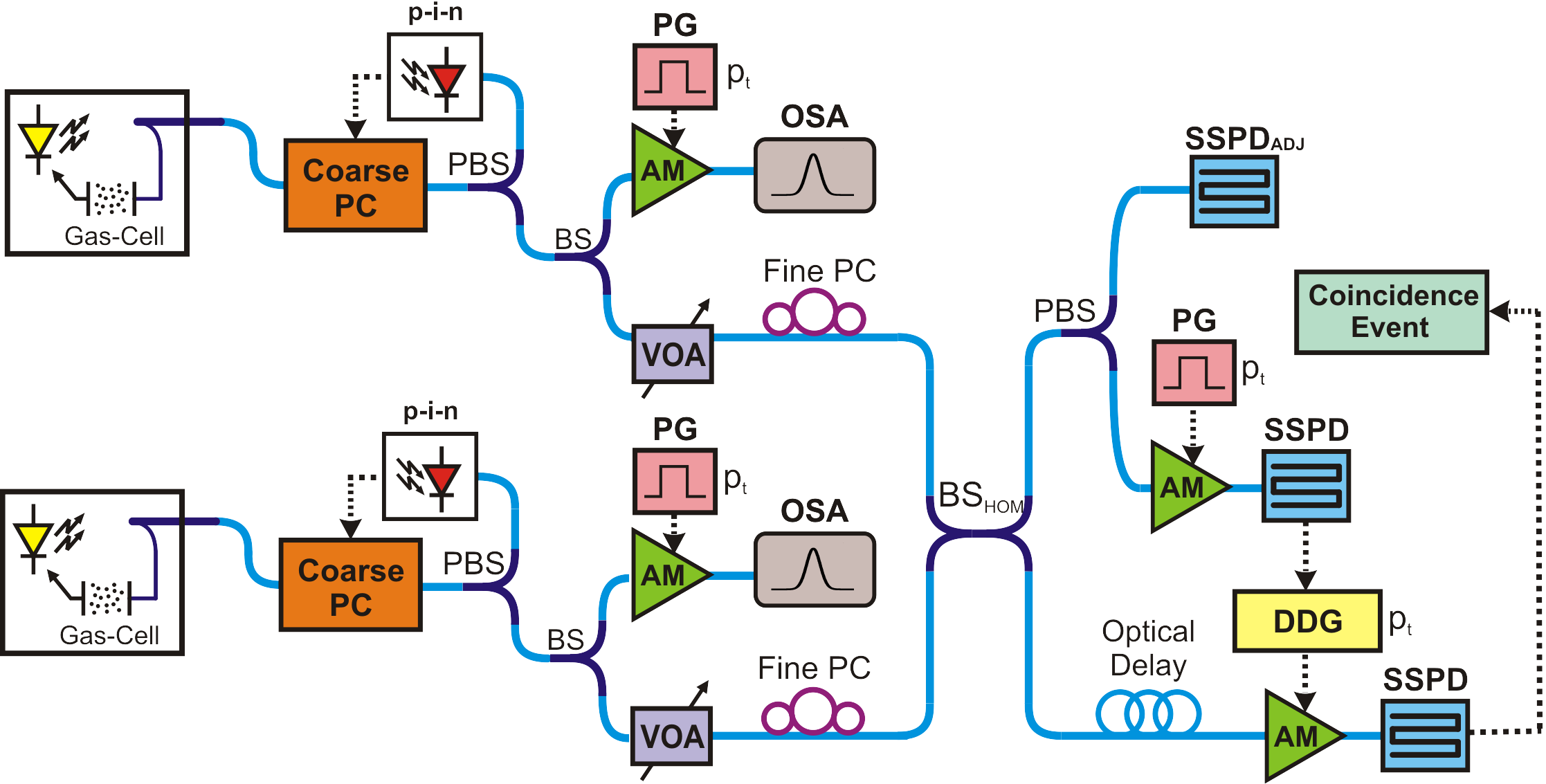}}
\caption{Block diagram of the experimental setup. The detections from SSPD slave are interpreted as the coincidence events and form the interferogram.}
\label{fig:Setup_SpectralWhichWay}
\end{figure}

Our experimental apparatus, presented in Fig. \ref{fig:Setup_SpectralWhichWay}, depicts the state preparation, interference, and measurement. The frequency-displaced coherent states are generated by two independent tunable wavelength lasers, which operate within the telecommunication band, and whose wavelength can be fine-tuned by a feedback signal. The lasers emission spectra have been adjusted to the absorption spectra of high-Q factor gas-cells through a PID system, which enables fine tuning of the order of MHz even within the absorption curve. The outputs of the frequency-tuned lasers are also polarization stabilized by active polarization controllers. Finally, variable optical attenuators and mechanical polarization controllers guarantee the fine-tuning necessary to guarantee indistinguishability in terms of polarization and intensity.

The prepared frequency-displaced WCSs are sent to the symmetrical beam-splitter BS$_{\textrm{HOM}}$. Connected to one of BS$_{\textrm{HOM}}$'s output is a polarizing beam-splitter (PBS) with one of the ports connected to a superconducting single-photon detector (SSPD$_{ADJ}$), which provides the feedback for fine-tuning the input polarizations with the mechanical polarization controllers; by minimizing the counts in this SSPD for both input states, one guarantees that the polarization states are aligned. The HOM interferogram is determined by placing two SSPDs -- master and slave -- at the remaining outputs of the PBS and of BS$_{\textrm{HOM}}$. Before both SSPDs (master and slave), however, an electro-optical amplitude modulator (AM) is introduced, which is responsible for chopping the optical pulse that arrives at the detector. The effect of the AMs is, thus, to emulate the detector gate since the SSPDs are free-running and have, ideally, infinite integration time, i.e., the AMs impose an imperfect frequency response to the detectors. An internally triggered pulse generator triggers the AM of SSPD master and, in the event of a detection, the AM of SSPD slave is also triggered after a tunable time-delay $\tau$; whenever coincidences from SSPD master and slave arrive at the same time at a coincidence unit, a coincidence count is registered; by sweeping the relative temporal delay $\tau$ between the pulses sent to AM master and AM slave the post-selected HOM interferogram is generated.

\begin{figure}[htbp]
\centering
\fbox{\includegraphics[width=0.77\linewidth]{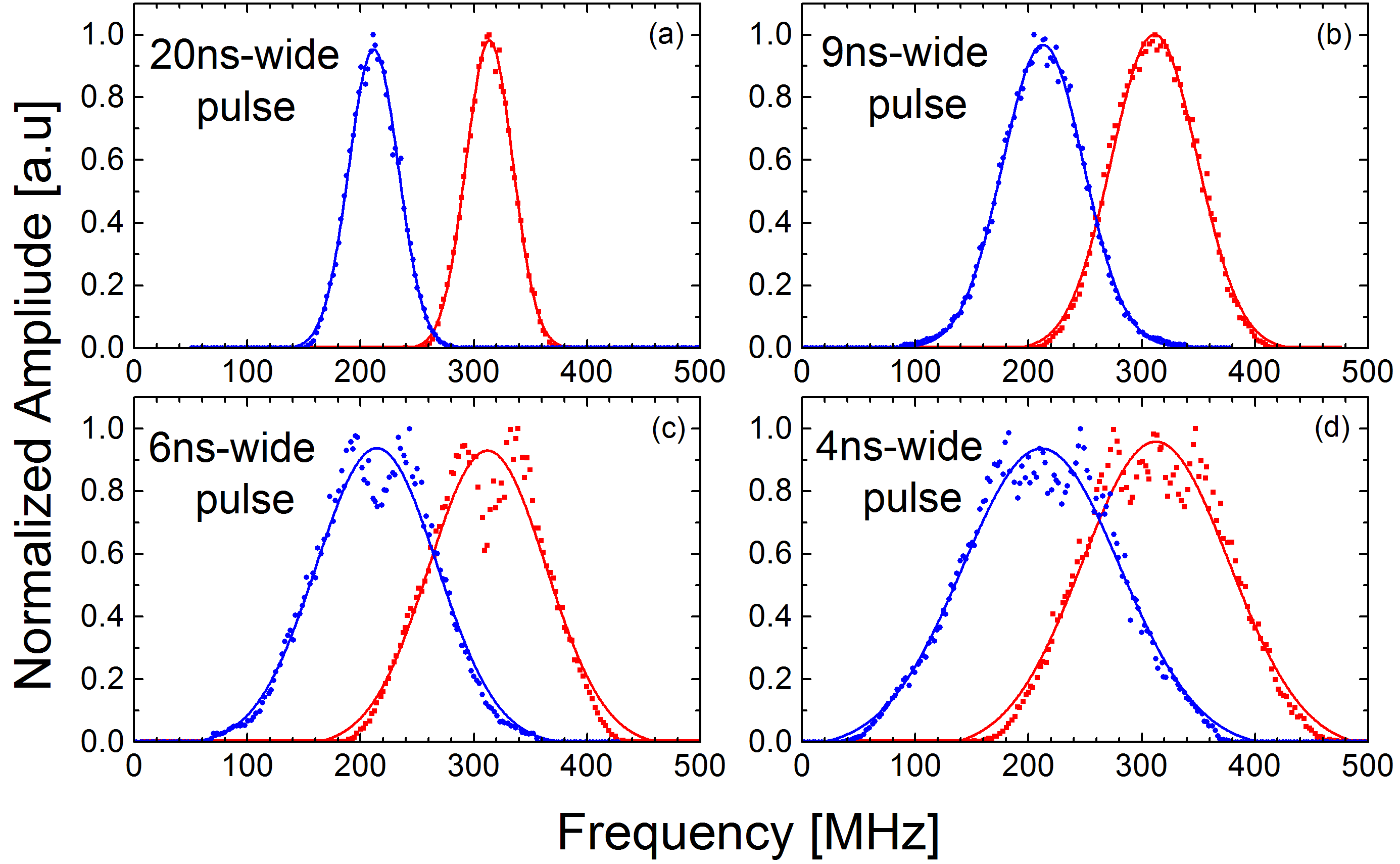}}
\caption{The spectral intensity distribution of the wave-packets for different values of the pulse width. The horizontal axis is shifted by 1551.3 nm. When $p_t$ is larger than 10 ns (a), the spectral overlap is minimum. When $p_t$ is narrower than 10 ns (c-d), however, the overlap grows significantly. Solid lines represent the gaussian fit.}
\label{fig:spectralOverlap}
\end{figure}

\begin{figure}[htbp]
\centering
\fbox{\includegraphics[width=0.77\linewidth]{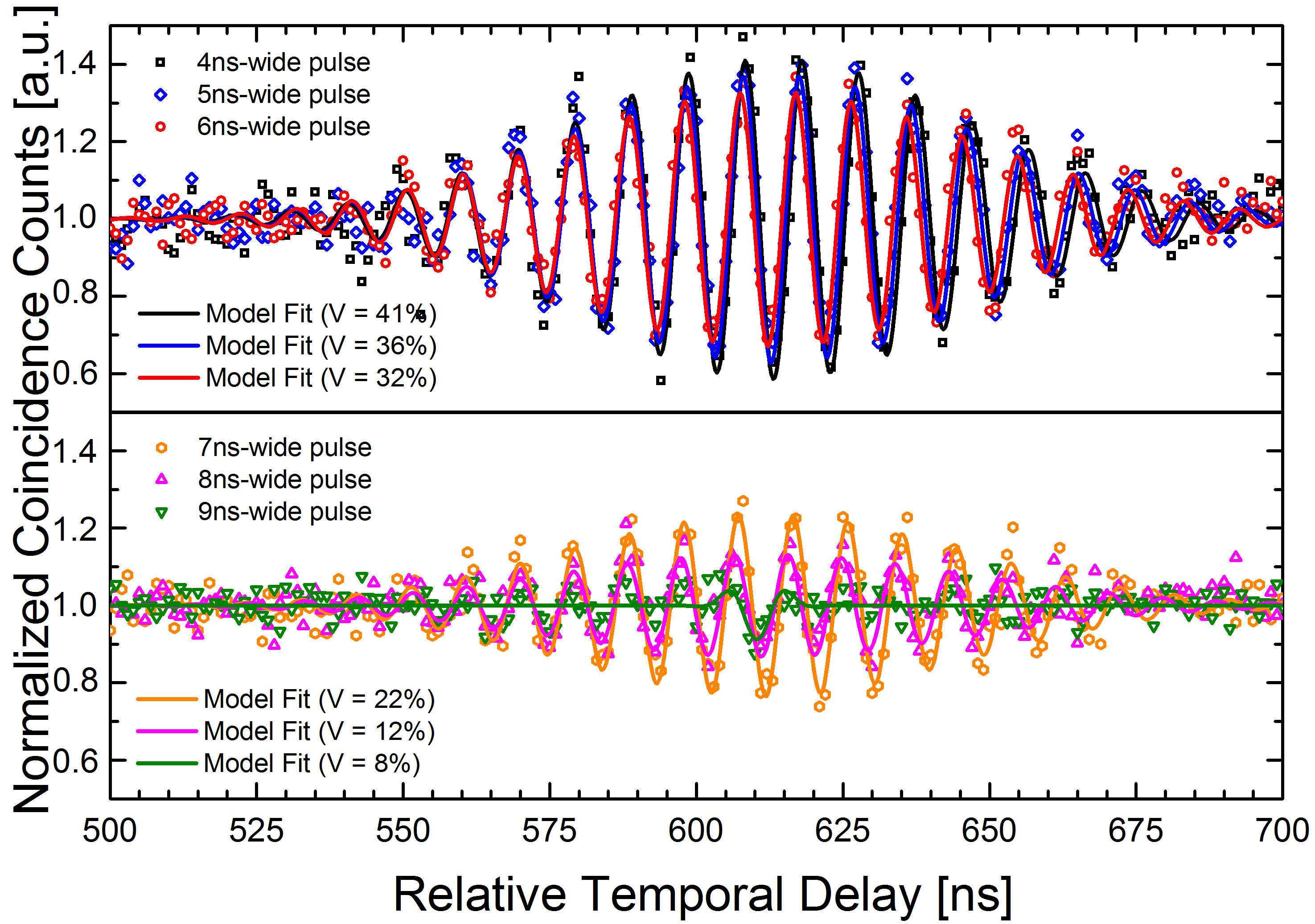}}
\caption{HOM interferograms acquired with different pulse widths. It is clear the visibility reduction as $p_t$ grows. Solid lines represent the model fit when gaussian shaped pulses are considered \cite{da2015linear}. As expected, the beat note is $\Delta \omega=100$ MHz.}
\label{fig:rawVis}
\end{figure}

For the experimental determination of the spectral complementarity relation between $K$ and $V$, the center frequencies of the laser sources are set to a fixed value so that their separation corresponds to 100 MHz. Since the spectral linewidth of each laser is on the order of 10 MHz ($\omega_{1,2}\pm 5$ MHz in the optical spectrum) and assuming that the trigger pulse sent to the AMs is larger than the corresponding coherence time of 100 ns, no interference pattern will be observed for these lasers, i.e., the photons will be absolutely distinguishable from the detector's perspective. In other words, upon measuring the wave-packets, the detectors are able to successfully identify their center frequency and, thus, determining their provenance, which causes no interference to take place -- recall that the wave-packets must be indistinguishable for HOM interference. By narrowing the AM trigger pulse widths ($p_t$) from 100 ns down to 4 ns (which is the minimum achievable by the employed pulse generator), an interference pattern can be observed when $p_t\leq10$ ns. Focusing, therefore, on the range between 10 to 4 ns, the effects of decreasing $p_t$ are two-fold and complementary: the spectral distributions are enlarged, since the chopping pulse is narrower than the optical source's temporal coherence, and end up overlapping, causing $K$ to decrease -- refer to Figs. \ref{fig:specWhichWay_Kvis} and \ref{fig:spectralOverlap}; meanwhile, the visibility of the HOM interferogram rises ($V$ rises) presenting a fixed 100 MHz interference fringe pattern -- refer to Fig. \ref{fig:rawVis}.

Even though the spectra are also enlarged when $p_t$ lies within the 100 to 10 ns region, it is not enough to create a non-zero spectral overlap, i.e., the product between $I_1\left(\omega\right)$ and $I_2\left(\omega\right)$ is always zero. With the righthand side of Eq. \ref{eq:intensityKnorm} zero, $K$ is unit, and, as expected from the proposal, $V$ is null, i.e., the visibility of the interferogram is zero. At $p_t\leq10$ ns, however, the linewidths are enlarged up to a point where a non-zero spectral overlap is already observed. Using the inverse relation between the linewidth and the coherence time $\tau_c=0.66/\Delta \omega$ \cite{saleh1991fundamentals} for gaussian wave-packets, we calculate the FWHM at $p_t=100$ ns is $\Delta \omega \approx 70$ MHz ($\omega_{1,2}\pm 35$ MHz), which is not enough for the distributions to overlap at half maximum ($\Delta \omega = \omega_1-\omega_2=100$ MHz), but enough so that the gaussian skirts overlap -- refer to Fig. \ref{fig:spectralOverlap} (b).

Before extracting the values of $K^2+V^2$ based on the experimental acquisitions, an important issue must be addressed: the achievable visibility is $50\%$ due to the multi-photon contribution of WCSs \cite{mandel1983photon} -- a detailed explanation is given in the Supplementary Information of \cite{jin2013two}. In order to distillate the multi-photon contribution from the coincidence count rates, a new visibility parameter is defined as
\begin{equation}
\mathcal{V} = \frac{\left(R_{dist}-0.5\cdot R_{dist}\right)-\left(R_{min}-0.5\cdot R_{dist}\right)}{\left(R_{dist}-0.5\cdot R_{dist}\right)}.
\end{equation}
$\mathcal{V}$ is calculated by subtracting the count floor imposed by the multi-photon pulses ($0.5\cdot R_{dist}$) and attempts to take into account the contributions from purely two-photon interference even when WCSs, in the few-photon regime, are employed. In other words, based on the fact that perfect indistinguishability between WCSs is given by $V=0.5$, the value of the visibility taking into account the quantum two-photon interference effect will be defined as $\mathcal{V}=V/0.5$.

\begin{figure}[htbp]
\centering
\fbox{\includegraphics[width=0.77\linewidth]{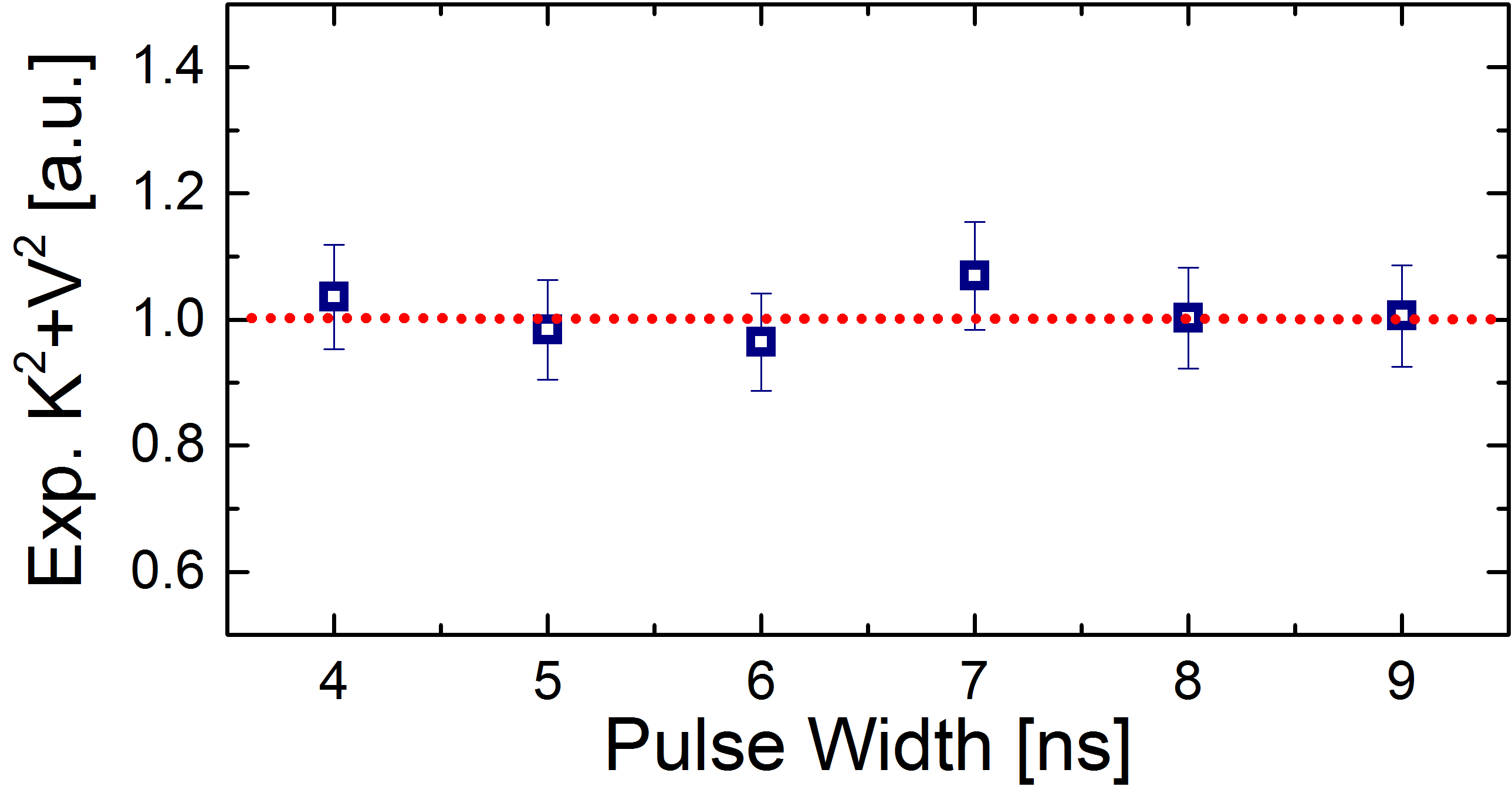}}
\caption{Experimental values of $K^2+\mathcal{V}^2$ traced against the pulse  $p_t$. The results follow the theoretical prediction -- represented by the dotted red line -- within the experimental error margin.}
\label{fig:complementarity}
\end{figure}

In Fig. \ref{fig:complementarity}, the experimental values of $K^2+\mathcal{V}^2$ are presented as a function of the width of the pulse $p_t$ sent to the AMs. The error bars correspond to fitting errors of the waveforms presented in Figs. \ref{fig:spectralOverlap} and \ref{fig:complementarity} from which the values of $K$ and $\mathcal{V}$ have been determined; first order approximations have been employed, for simplicity, in order to account for the non-linear association between $K$ and $\mathcal{V}$.

In conclusion, the distinguishability between photonic wave-packets has been discussed in the spectral context, focusing on the two-photon interference phenomenon in a Hong-Ou-Mandel interferometer fed with weak-coherent states. An experimental and theoretically sound definition of the spectral distinguishability parameter $K$ has been set forward based on the fidelity between quantum states. The complementarity relation between $K$ and a consistently defined visibility parameter $\mathcal{V}$ has been observed with weak-coherent states. The results attest that $K$ and $\mathcal{V}$ hold a complementarity relation, i.e., even though a spectral measurement is not performed by the single-photon detectors, the mere possibility of assessing the spectral distinguishability information is enough to impact on the visibility of the resulting interferogram. Furthermore, the quantum aspect of the two-photon interference in a Hong-Ou-Mandel interferometer could be examined with weak-coherent states.

\section*{Acknowledgements}

G. C. Amaral is indebted to Dr. K. Lemr for invaluable discussions and comments. The authors acknowledge the financial support of brazilian agencies CNPq, CAPES, and FAPERJ.


\begin{thebibliography}{10}

\bibitem{hong1987measurement}
C.~Hong, Z.-Y. Ou, and L.~Mandel, \emph{Physical Review Letters}, vol.~59,
  no.~18, p. 2044, 1987.

\bibitem{jin2013two}
J.~Jin, J.~A. Slater, E.~Saglamyurek, N.~Sinclair, M.~George, R.~Ricken,
  D.~Oblak, W.~Sohler, and W.~Tittel, ``Two-photon interference of weak
  coherent laser pulses recalled from separate solid-state quantum memories,''
  \emph{Nature communications}, vol.~4, 2013.

\bibitem{lo2012measurement}
H.-K. Lo, M.~Curty, and B.~Qi, \emph{Physical review letters}, vol. 108,
  no.~13, p. 130503, 2012.

\bibitem{gisin2002quantum}
N.~Gisin, G.~Ribordy, W.~Tittel, and H.~Zbinden, ``Quantum cryptography,''
  \emph{Reviews of modern physics}, vol.~74, no.~1, p. 145, 2002.

\bibitem{mandel1983photon}
L.~Mandel, \emph{Physical Review A}, vol.~28, no.~2, p. 929, 1983.

\bibitem{da2015linear}
T.~F. da~Silva, G.~C. Amaral, G.~P. Tempor{\~a}o, and J.~P. von~der Weid,
  ``Linear-optic heralded photon source,'' \emph{Physical Review A}, vol.~92,
  no.~3, p. 033855, 2015.

\bibitem{amaral2016few}
G.~C. Amaral, T.~F. da~Silva, G.~P. Tempor{\~a}o, and J.~P. von~der Weid,
  ``Few-photon heterodyne spectroscopy,'' \emph{Optics letters}, vol.~41,
  no.~7, pp. 1502--1505, 2016.

\bibitem{da2013proof}
T.~F. da~Silva, D.~Vitoreti, G.~Xavier, G.~do~Amaral, G.~Temporao, and
  J.~von~der Weid, ``Proof-of-principle demonstration of
  measurement-device-independent quantum key distribution using polarization
  qubits,'' \emph{Physical Review A}, vol.~88, no.~5, p. 052303, 2013.

\bibitem{legero2003time}
T.~Legero, T.~Wilk, A.~Kuhn, and G.~Rempe, ``Time-resolved two-photon quantum
  interference,'' \emph{Applied Physics B: Lasers and Optics}, vol.~77, no.~8,
  pp. 797--802, 2003.

\bibitem{vedral2006introduction}
V.~Vedral.\hskip 1em plus 0.5em minus 0.4em\relax Oxford University Press on
  Demand, 2006.

\bibitem{saleh1991fundamentals}
B.~E. Saleh, M.~C. Teich, and B.~E. Saleh.\hskip 1em plus 0.5em minus
  0.4em\relax Wiley New York, 1991, vol.~22.

\bibitem{lewis1995fast}
J.~Lewis, in \emph{Vision interface}, vol.~10, no.~1, 1995, pp. 120--123.

\bibitem{englert1996fringe}
B.-G. Englert, \emph{Physical review letters}, vol.~77, no.~11, p. 2154, 1996.

\end{thebibliography}
\end{document}